\documentclass[12pt,english,twoside, reqno]{article}

\usepackage{graphicx}
\usepackage{amssymb}
\usepackage{textcomp}

\usepackage{cite}

\usepackage{bm}

\usepackage{mathabx}

\usepackage{dingbat}

\usepackage{pifont}

\usepackage{fancyhdr}
\usepackage[english,french]{babel}
\usepackage[latin1]{inputenc}
\usepackage{tipa}
\usepackage{bbm}

\usepackage{graphicx}
\usepackage{amssymb}
\usepackage{textcomp}
\usepackage{stmaryrd}

\begin{document}

\newcommand{\nin}{\noindent}
\newcommand{\vv}{\vskip 0.25 cm}
\newcommand{\vvv}{\vskip 0.3 cm}
\newcommand{\vvvv}{\vskip 0.5 cm}

\newcommand{\beq}{\begin{equation}}
\newcommand{\enq}{\end{equation}}
\newcommand{\und}{\underline}
\newcommand{\biz}{\begin{itemize}}
\newcommand{\eiz}{\end{itemize}}
\newcommand{\di}{\displaystyle}
\newcommand{\bc}{\begin{center}}
\newcommand{\ec}{\end{center}}

\newcommand{\pa}{\partial}
\newcommand{\g}{\gamma}
\newcommand{\ep}{\epsilon}

\newcommand{\ov}{\overline}

\newcommand{\vp}{\varphi}

\newcommand{\cip}{{\bms P}}
\newcommand{\ciq}{{\bms Q}}
\newcommand{\op}{{\cal P}}
\newcommand{\oec}{{\cal E}_c}

\newcommand{\opx}{{\cal P}_x}
\newcommand{\opy}{{\cal P}_y}
\newcommand{\olz}{{\cal L}_z}

\newcommand{\zb}{\bar{z}}
\newcommand{\zbo}{{\bar{z}}_0}

\newcommand{\av}{\cal V}

\newcommand{\bms}{\boldsymbol}


\vskip 1.5 true cm

\centerline{\bf \large A note on Gurzadyan's theorem}

\vskip 0.5 true cm 

\vskip 0.75 cm
\centerline{{\bf Christian CARIMALO}\footnote{{\bf christian.carimalo@sorbonne-universite.fr}}}

\vskip 0.75 cm
\centerline{Sorbonne Universit\'e, Campus Pierre et Marie Curie}
\centerline{4 Place Jussieu, 75005 Paris, France}

\vskip 1.5cm
\centerline{ABSTRACT}

\vvv \nin  The issue and proof of Gurzadyan's theorem are presented concisely, avoiding tedious and unnecessary calculations that would mask what is essential. The goal is to provide a good mathematical and physical understanding of the theorem, making you want to learn more about its use in cosmology.  

\vskip 0.75 cm

\noindent {\bf Keywords} : Classical mechanics, Gravitation, Cosmological Constant.
 
\vskip 0.75cm

\counterwithin{equation}{section}

\section{\large Newton's shell theorem} 

\vv \nin We will start with a presentation of Newton's famous shell theorem, of which Gurzadyan's theorem is an extension. 

\vv \nin We consider masses uniformly distributed on a sphere $S$ at an areal density of $m$. The sphere is centered on point $C$ and has radius $b$. From the superposition principle, the overall gravitational potential and gravitational field at point $P$ are, respectively,   

$$ U (P,C) = \di{\int_{S(C,b)}}   d U(P,Q),~~{\rm where}~~ dU(P,Q) = m\, dS(Q) \, u(P,Q) ~~$$  
\beq {\rm and} ~~{\bms G}(P,C) = - \nabla_P U(P,C)  \label{VG} \enq 

\vv \nin $u(P,Q)$ being a function of both positions of point $P$ and point $Q$, the latter being on the sphere, and given by $u(P,Q) = G/(PQ)$ from Newton's law of gravity ($G$ is the gravitational constant). Using Gauss's theorem and considering the spherical symmetry of the distribution, it is easy to find the following result, first published by Newton in 1687 \cite{Newt}, and referred to as  Newton's shell theorem : 

\vv \nin $\bullet$ The distribution acts on a test mass at a point $P$ outside the sphere as if  
its mass $M = 4 \pi b^2 m$ was concentrated at its center $C$, that is, ${\bms G}(P,C) = -G M \di{{\bms {CP}}\over{CP^3}} $ ;   
 
\vv \nin $\bullet$ the gravitational field is zero at any point $P$ (strictly) inside the sphere. 

\vv \nin As done in Ref. \cite{Reed}, it is relevant to ask whether this result is a direct consequence of Newton's gravitational law (the inverse square law for the field). Without presupposing the form of $u(P,Q)$, we write the potential as :    

\beq U(P,C) = M\, \bar{u}_P(C),~~{\rm with}~~ \bar{u}_P(C) = \di{1 \over{4 \pi b^2}} \di{\int_{S(C,b)}} dS(Q) \, u(P,Q) \label{mean1} \enq  

\vv \nin Defining $u_P(Q) = u(P,Q)$, it becomes apparent that $\bar{u}_P(C)$ is the mean value of the function $u_P(Q)$ at point $C$, when this function is integrated over the sphere $S(C,b)$. Under what conditions would we have $\bar{u}_P(C) = u_P(C)$ ? This happens in two particular cases. One is trivial : when $u(P,Q)$ is constant. The other appears when $PC \gg b$, because the distribution is then seen from point $P$ as a point mass $M$ located at $C$. However, this does not provide any information regarding the law of forces. Is there a solution for any finite $b$ ? The answer is given by a well-known mathematical theorem (the ``Mean value theorem", see e.g. Ref. \cite{ABR}) stating that the equality $\bar{u}_P(C) = u_P(C)$ is an exclusive property of harmonic functions : 

\beq   \bar{u}_P(C) = u_P(C)~~{\rm if~and~only~if} ~~\Delta_C\, u_P(C) =0 \label{mean2} \enq   

\vv \nin $\Delta_C$ being the Laplacian operator with respect to the coordinates of point $C$. From the physicist's perspective, we assume that all mathematical operations necessary for a physical description are allowed. In appendix B is given a presentation of this theorem for an euclidean space of dimension $n \geq 3$, using the appropriate Green function and Dirac distribution. 

\vv \nin If we admit the principle that interactions are central at the corpuscular level, from which it follows that $ u(P,C)$ depends on the positions of $P$ and $C$ only through their relative distance $a=PC$, i.e. $ u(P,C) = u(a)$, then 

\beq \Delta_C \,u(P,C) = \Delta_P \,u(P,C) =  \di{1\over a} \di{d^{2}\over {d a^{2}}} \left[a u(a)\right] = 0  \label{lap} \enq

\vv \nin Hence, the unique admissible form $u(a) = \di{K_1 \over a} +K_2$, $K_1$ and $K_2$ being constants, and the inverse-square law for the gravitational field, at least for $a \geq b$. 
Note that $u(a)$ being harmonic, the mean property stated above does not depend on the value of $b \leq a$, more precisely, the constants $K_1$ and $K_2$ corresponding to the outside of the sphere do not depend on $b$. The first part of Newton's shell theorem is thus fully explained. As will be seen now, the second part of this theorem is more subtly related to Newton's gravitational law.  

\vv \nin Assuming $u(P,Q) = u(PQ)$, the mean value in Eq. (\ref{mean1}) is expressed as    

\beq \bar{u}_P(C) = I(a,b) = \di{1 \over 2} \di{\int^1_{-1}} dx \, u(r),~~~{\rm with}~~~r = \sqrt{a^2 +b^2 +2 ab x} \label{Iab} \enq 

\vv \nin Due to the symmetry $I(a,b) = I(b,a)$, it is obvious that if $I(a,b)$ does not depend on $b$ when $a \geq b$, it cannot depend on $a$ when $b \geq a$, whatever the form of $u(r)$. Consequently, inside the sphere, the potential is constant and the field is zero. Only the fact that $\bar{u}_P(C)$ does not depend on $b$ when $b \leq a$ is due to the inverse-square law, as seen previously. The spherical symmetry of the distribution is the final crucial point leading to the conclusion.   

\vv \nin At this stage, it is tempting to check the result in Eq. (\ref{lap}). Taking $s = r^2$ as the integration variable, Eq. (\ref{Iab}) is rewritten as 

\beq I(a,b) = u(a) = \di{1\over{4 ab}} \di{\int^{s_2}_{s_1}} ds\, u(\sqrt{s} ), ~~{\rm where}~~s_2= (a+b)^2,~s_1=(a-b)^2, \enq  

\vv \nin and for $a \geq b$ we have 

$$ \di{\partial \over{\partial b}} \left[ 2 ab u(a) \right] = 2 a u(a) = \di{1 \over 2} \left[ u(\sqrt{s_2})  \di{{\partial s_2}\over{\partial b}} -  u(\sqrt{s_1})  \di{{\partial s_1}\over{\partial b}} \right] $$ 
$$ =   (a+b) u(a+b) + (a-b) u(a-b) , ~~{\rm or} $$ 
\beq \left[ (a+b) u(a+b) - a u(a) \right] +  \left[ (a-b) u(a-b) - a u(a) \right] =0 \label{zero} \enq

\vv \nin which equation must be satisfied whatever the value of $b$. Expanding in powers of $b$ its left member which is symmetrical with respect to the change of $b$ into $-b$, we find    

\beq \di{\sum^\infty_{k=1}}  \di{b^{2k} \over{(2k)!}} \di{d^{2k}\over {d a^{2k}}} \left[a u (a)\right]  = 0 \label{fin1} \enq

\vv \nin and, as expected, this is realized if and only if $\Delta u  = \di{1\over a} \di{d^{2}\over {d a^{2}}} \left[a u (a)\right] = 0$.

\section{\large Gurzadyan's theorem} 

\vv \nin Gurzadyan's theorem is an extension of the first conclusion of Newton's shell theorem, answering the question : what is the more general law of force such that the spherical symmetric mass distribution on $S(C,b)$ acts on a test mass outside the sphere exactly as a point mass $M$ located at its center $C$ ? 

\vv \nin To our knowledge, this theorem has so far only received a truncated proof by Gurzadyan himself \cite{Gurz}, or proofs based either on power-series methods \cite{Reed, kiki1}, or on a ``perturbative method", according to the author of Ref. \cite{TVP}, or in a way that does not highlight the essential points \cite{mark}. In the following we propose a concise and clear proof rid of unnecessary calculations.   

\vv \nin  The novelty in the hypothesis is that the equality  $\bar{u}_P(C) = u(P,C)$ is no longer required but replaced by  

\beq {\bms G}(P,C) = - M \nabla_P\, \bar{u}_P(C) = - M \nabla_P \,u(P,C) ~~{\rm outside~the~ sphere} \label{gur1} \enq 

\vv \nin This condition is satisfied in the three following situations : $u=$ constant,    
$u$ is harmonic, and when $a \gg b$. For finite $b/a$, we will proceed as follows.  

\vv \nin Thanks to the spherical symmetry of the distribution, this condition is simplified into  

\beq \di{ \partial \over{\partial a}} \left[ \bar{u}(a,b) - u(a) \right] =0  \label{gur2} \enq

\vv \nin for $ a \geq b$ and {\it whatever the value of $b$}. Note that here $u(P,C)=u(a)$ is supposed to be independent of $b$ while the mean value $\bar{u}(a,b)$ can depend on both $a$ and $b$. Let us write 

$$ X= \di{ \partial \over{\partial b}} \left[ b \di{ \partial \over{\partial a}} \left( \bar{u}(a,b) - u(a) \right) \right] =\di{ \partial \over{\partial a}} \di{1 \over a} \di{ \partial \over{\partial b}} \left[ a b \left( \bar{u}(a,b) - u(a) \right) \right]= 0 $$ 

\vv \nin Then, repeating the calculation leading to Eq. (\ref{fin1}), we find 

$$ X = \di{ \partial \over{\partial a}} \di{1 \over a} \left[ \di{\sum^\infty_{k=1}}  \di{b^{2k} \over{(2k)!}} \di{d^{2k}\over {d a^{2k}}} \left[a u (a)\right] \right] 
= \di{\sum^\infty_{k=1}}  \di{b^{2k} \over{(2k)!}} \di{d^{2k-1}\over {d a^{2k-1}}} \Delta u(a) $$ 
\beq = \di{b^2 \over 2}  \di{ d \over{d a}} \Delta u (a)+ \cdots \enq

\vv \nin and conclude that Eq. (\ref{gur1}) (i.e. $X=0$) is realized if and only if 

\beq \di{ d \over{d a}} \Delta u(a) =  \di{ d \over{d a}} \left[ \di{1 \over a} \di{{d^2} \over{da^2}} \left[a u(a)\right] \right] = 0 \label{gurz3} \enq  

\vv \nin Expressing the field as ${\bms G}(P,C) = - M g(a) \di{{\bms {CP}}\over a} $ with $ g(a) = - \di{{du}\over{da}}(a)$, Eq. (\ref{gurz3}) written in terms of $g(a)$ and its derivatives with respect to $a$ is     

\beq  \di{a^2 \over 2} g^{\prime \prime} + a g^\prime -  g =0  \enq

\vv \nin which is identical to Eq. 4 of Ref. \cite{Gurz}. Integration of Eq. (\ref{gurz3}) gives the general expression of admissible functions $u(a)$ : 

\beq u(a) = \di{K_1 \over a} + K_2 + \di{{K_3 a^2}\over 2}  \label{gurz5} \enq 

\vv \nin $K_1$, $K_2$ and $K_3$ being arbitrary contants. This is the final answer of Gurzadyan's theorem : besides the Newtonian field $\propto \,1/a^2$, the only field possessing the property required is of the Hookean type $\propto\, a$. 

\vv \nin Note that taking $u(P,Q) = r^2$ yields $\bar{u}(P,C) = a^2 + b^2$, depending on both $a$ and $b$, which reflects the fact that the strict equality $\bar{u}(P,C) = u(P,C) (= a^2)$ cannot be realized if $u$ is not harmonic. However, the relation $\nabla \bar{u}(P,C) = \nabla u(P,C)$ holds for the Hookean function, but its field does not vanish inside the sphere, except at point $C$.    

\vv \nin From the physical point of view, it is important to remark that the Hookean field cannot be considered as a classical gravitational field or a Coulomb field, because in a region empty of sources the divergences of the latter are commonly considered to be zero, whereas the divergence of a Hookean field is never zero ($\Delta r^2 = 6$). In cosmology, Gurzadyan considered the Hookean field as a correction to the traditional gravitational field, that could explain in a natural way the appearance of the cosmological constant in Einstein equation of General Relativity  \cite{gurz2,gurz3}. However, in Classical Physics, the divergences of fields are usually considered as cause and effect relationships connecting fields to their sources. According to this, the sources of the Hookean field would not only be located on the surface $S$, but would actually fill the entire space uniformly. Then arises the question of the meaning of this strange Hookean field itself, or, if the previous interpretation is incorrect, that of the true relationship between the fields and their sources.  

\vv \nin In the general case, the mean value $\bar{u}(P,C)$ does depend on both $a$ and $b$ and it seems difficult to achieve an extension of Gurzadyan's theorem. However, various authors have considered the interesting case where this mean value appears in the form of the product $ u(a) w(b)$ \cite{Chap, BK,Raut}. The result is given in appendix A.   

\newpage

\bibliographystyle{amsplain}

\renewcommand{\refname}{\large{References}}

\newpage


\nin {\large \bf Appendix A : An extension of Gurzadyan's theorem } \label{ApG} 

\setcounter{equation}{0}
\renewcommand{\theequation}{\mbox{A.}\arabic{equation}}

\vv \vv \nin Let's look for what types of functions $u(PQ)$ and $w(b)$ could satisfy the relation 

\beq \bar{u}(P,C) = u(a) \, w(b)  = \di{1\over{4 ab}} \di{\int^{s_2}_{s_1}} ds\, u(\sqrt{s} ) \enq   

\vv \nin We have 

$$  \di{\partial \over{\partial b}} \left[ 2 ab u(a) w(b) \right] = 2 a u(a)  \di{d \over{d b}}  \left[ b w(b) \right] = \di{1 \over 2} \left[ u(\sqrt{s_2})  \di{{\partial s_2}\over{\partial b}} -  u(\sqrt{s_1})  \di{{\partial s_1}\over{\partial b}} \right] $$ 

\beq = \left[ (a+b) u(a+b) - a u(a) \right] +  \left[ (a-b) u(a-b) - a u(a) \right] + 2 a u(a) \enq

\vv \nin Hence, 

\beq  a u(a) \left[ \di{d \over{d b}}  \left[ b w(b) \right] -1 \right] = \di{\sum^\infty_{k=1}}  \di{b^{2k} \over{(2k)!}} \di{d^{2k}\over {d a^{2k}}} \left[a u (a)\right]  \label{EG1} \enq

\vv \nin It is obvious that this equation is satisfied if and only if $\di{d^{2k}\over {d a^{2k}}} \left[a u (a)\right] = c_k \left[a u (a)\right] $ where $c_k$ is constant. Thus, we must have 

$$ \di{d^{2}\over {d a^{2}}} \left[a u (a)\right] = c_2 \left[a u (a)\right],$$
$$ \di{d^{4}\over {d a^{4}}} \left[a u (a)\right] = c_4 \left[a u (a)\right] = \di{d^{2}\over {d a^{2}}} \left[ \di{d^{2}\over {d a^{2}}} \left[a u (a)\right] \right] = c^2_2 \left[a u (a)\right], $$ 
$$ \di{d^{6}\over {d a^{6}}} \left[a u (a)\right] = c^3_2 \left[a u (a)\right],~~{\rm etc.}$$

\vv \nin and finally, Eq. (\ref{EG1}) is realized only by functions $u$ and $w$ verifying  

 \beq \di{d^{2}\over {d a^{2}}} \left[a u (a)\right] = c_2 \left[a u (a)\right]~~{\rm and}~~\di{d \over{d b}}  \left[ b w(b) \right] -1 = \di{\sum^\infty_{k=1}}  \di{ {c^k_2\, b^{2k}} \over{(2k)!}} \enq

\vv \nin Integrating these equations gives the following forms. 

\vv \nin $\bullet$ If $c_2 =0$, 

\beq u(a) = \di{K_1 \over a} + K_2, ~~w(b) = 1 + \di{K_3 \over b} \enq   

\vv \nin $\bullet$ If $c_2 >0$, we set $c_2 = \lambda^2$. Then, 

$$  \di{d \over{d b}}  \left[ b w(b) \right] =1 + \di{\sum^\infty_{k=1}}  \di{ {(\lambda b)^{2k}} \over{(2k)!}} = \cosh \lambda b, ~~{\rm and} ~~w(b) = \di{{\sinh \lambda b}\over {\lambda b}} +  \di{K_3 \over b}, $$
\beq u(a) = K_1 \di{{\exp(- \lambda a)}\over a} + K_2  \di{{\exp( \lambda a)}\over a} \enq
 
\vv \nin $\bullet$ If $c_2 < 0$, we set $c_2 = -\omega^2$. Then, 

$$  \di{d \over{d b}}  \left[ b w(b) \right] =1 + \di{\sum^\infty_{k=1}} (-1)^k \di{ {(\omega b)^{2k}} \over{(2k)!}} = \cos \omega b, ~~{\rm and} ~~w(b) = \di{{\sin \omega b}\over {\omega b}} +  \di{K_3 \over b}, $$
\beq u(a) = K_1 \di{{\cos(\omega a)}\over a} + K_2  \di{{\sin( \omega a)}\over a} \enq

\vv \nin In each case, $K_1$, $K_2$ and $K_3$ are constants with appropriate dimensions. 

\vv \nin The physical content of this extension is that the field at a point $P$ outside the sphere is the same as that of a point source located at the center $C$, the mass of which is no longer the total mass $M$ of the distribution, but the product of the latter by a dimensionless factor $w(b)$ depending on the radius of the sphere : ${\bms G}(P,C) = -G M' \nabla_P\, u(P,C)$ with $M' = M w(b)$. Among non-harmonic potentials satisfying this property, the Yukawa potential $u(r) = e^{-\lambda r}/r$ ($\lambda > 0$) shows a good behavior for physical modelisations. In fact, it has a long history in Physics, particularly in Particle Physics \cite{row}. Nowadays, it often appaears in this form or modified forms in research works on gravitation theory \cite{GN,wis,beck}, and the phenomenology of dark universe 
\cite{AJCL,JLM}.

\newpage

\nin{\large \bf Appendix B : Extension to $n$ dimensions }  

\setcounter{equation}{0}
\renewcommand{\theequation}{\mbox{B.}\arabic{equation}}

\vv \vv \nin In an euclidean space of dimension $n \geq 3$, the equivalent of Newton's potential $1/r$ in three dimensions is $U_N(r)= 1/ r^{n-2}$, where $r = \sqrt{x^2_1 + x^2_2 + \cdots +x^2_n}$. It satisfies the equation 

\beq \Delta U_N = - k_n \,\delta({\bms r}),~~~{\rm where}   \label{del} \enq  

\vv \nin $\bullet$ $\Delta$ is the Laplace operator $ \Delta = \di{\sum^n_{i=1}} \di{\partial^2\over{\partial x^2_i}} $, which, when applied to a spherical symmetric function,  reduces to    

$$ \Delta = \di{{d^2}\over {d r^2}} + \di{{n-1}\over r} \di{{d}\over {d r}} $$

\vv \nin $\bullet$ $\delta({\bms r}) = \di{\prod^n_{i=1}}\,  \delta(x_i)$ is the Dirac distribution for dimension $n$, such that, for any volume $V$, any point $C$, and any sufficiently regular function $F$  

\beq \di{\int_V}\,\delta({\bms r}_Q - {\bms r}_C)\, F(Q) \,dV(Q) = \theta_V(C) \, F(C) \label{fdel} \enq  

\vv \nin where $\theta_V(C) =1$ if $C$ is inside $V$ and  $\theta_V(C) =0$ if $C$ is outside $V$ ; 

\vv \nin $\bullet$ $k_n = (n-2) \Omega_n$ where $\Omega_n = 2 \di{\pi^{n/2} \over \Gamma(n/2)}$ is the total solid angle, $\Gamma$ being the Euler's Gamma function. 

\vv \nin Considering a ball $B$ of volume $V$ delimited by the sphere $S$ of radius $b$ and centered at $C$, Green's theorem states that for any functions (or even distributions) $u$ and $v$, we have   

\beq \di{\int_S} \left[ v \nabla u - u \nabla v \right] \cdot d {\bms S} = \di{\int_B} \left[ v \Delta u - u \Delta v \right] d V \label{green} \enq 

\vv \nin Let's take $v = (CQ)^{2-n} - b^{2-n}$ for which $ v=0$ when $Q$ is on the surface $S$ and $\Delta_Q \, v = - k_n \delta({\bms {CQ}})$. We have   

$$  \di{\int_S} \left[ v \nabla u - u \nabla v \right] \cdot d {\bms S} =  b^{1-n} (n-2)  \di{\int_S}  u  d S  $$
$$ = \di{\int_B} \,\left[r^{2-n} - b^{2-n} \right]  \Delta u \, d V +  k_n \,u(P,C) $$

\vv \nin Since $d S = b^{n-1} d \Omega_n$ and $\bar{u}(P,C) = \di{1\over \Omega_n} \int \, d\Omega_n \, u(P,Q) $ being the mean value of $u$ at point $C$, we obtain 

\beq \bar{u}(P,C) - u(P,C) = \di{1\over k_n} \di{\int_B} \left[ r^{2-n} - b^{2-n} \right] \Delta_Q\, u (P,Q)\, d V(Q)  \label{gap} \enq

\vv \nin with $dV = r^{n-1} dr d \Omega_n$. It is then obvious that $\bar{u}(P,C) = u(P,C)$ if $u$ is harmonic, i.e. if $\Delta_Q\, u(P,Q) =0 $ everywhere in the volume $V$ (note that the factor $r^{2-n} - b^{2-n}$ prevent singularities when $r=b$).   

\vv \nin Conversely, suppose that $\bar{u}(P,C) = u(P,C)$ {\it independently of} $b$. Then, taking $b$ as small enough, we can write 

$$0 = \di{1\over k_n} \di{\int_B} \left[ r^{2-n} - b^{2-n} \right] \Delta u \, d V \approx \di{1 \over k_n} \Delta_C u(P,C) \,\Omega_n \di{\int^b_0} \left[ r - r^{n-1} b^{2-n} \right] dr $$
$$ = \di{b^2 \over{2n}} \Delta_C \,u(P,C) $$
 
 \vv \nin and therefore $\Delta_C\, u(P,C) = 0$, which means that $u$ must be harmonic.  Considering here central functions, i.e. functions such that $u(P,Q) = u(PQ)$, we have as well 
$\Delta_P\, u(P,C) = \Delta_C\, u(P,C) =0$. The approximate relation 
 
 \beq   \bar{u}(C) - u(C) \approx \alpha \, b^2\, \Delta_C\, u(C) \label{anloc} \enq 
 
 \vv \nin for any function $u(C)$ provides an interpretation of its Laplacian as a measure of the {\it local anomaly} of that function i.e. the gap between the value of a function at a given point and the average of its values in the vicinity of this point, see e.g. \cite{mcdo,siv,styer}. The numerical constant $\alpha$ is a priori specific to the closed surface surrounding $C$ and $b$ is a characteristic dimension of this surface.  
 

\vv \nin More generally, the second member of Eq. (\ref{gap}) can be expressed as a series in powers of $b$, \cite{simi, olev}. First, for central potential $u(PQ)$, let's rewrite Eq. (\ref{gap}) in the form 

$$ \bar{u}(P,C) - u(PC) = \di{1\over k_n}  \Delta_P \, \di{\int} d \Omega_n  \di{\int^b_0} \vp_0 (r) \, u(PQ) \, dr ~~{\rm with}  $$ 
\beq  \vp_0(r) = r - r^{n-1} b^{2-n}  \enq 
 
\vv \nin Then, let's expand $u(PQ)$ in the Taylor series 

$$ u(PQ) = u(PC) + \di{\sum^\infty_{k=1}} \, \di{r^k\over {k!}} \di{\sum^n_{i_p=1}} \hat{x}_{i_1} \cdots \hat{x}_{i_k} \partial_{i_1} \cdots \partial_{i_k} u (PC) $$ 

\vv \nin where $x_i$ are the cartesian coordinates of $Q$ relative to $C$, $r = CQ$, $\hat{x}_i = x_i/r $ and $\partial_i = \partial/\partial x_i$. Thus, 

$$ \bar{u}(P,C) - u(PC) =  \Delta_P \di{\sum^\infty_{k=0}} \, I_k \,\di{\sum^n_{i_p=1}}\, T_{i_1\cdots i_k} \, \partial_{i_1} \cdots \partial_{i_k} u (PC) $$
$$ {\rm where}~~~I_k =  \di{1\over{n-2}}  \, \di{1\over {k!}} \,\di{\int^b_0} r^k \vp_0(r) dr  ~~~{\rm and} $$
$$  T_{i_1\cdots i_k} = <\hat{x}_{i_1} \cdots \hat{x}_{i_k} > = \di{1\over \Omega_n} \di{\int} d \Omega_n \,
\hat{x}_{i_1} \cdots \hat{x}_{i_k} $$

\vv \nin Note that, mathematically speaking, $\vp_0(r)$ is identically zero if $n=2$. Consequently, the integral in $I_k$ must be proportional to $ n-2$, thus cancelling the factor $1/(n-2)$ ; it is  also proportionnal to $b^{k+2}$. Due to the spherical symmetry, the tensor $T_{i_1\cdots i_k}$ is non-zero only if $k$ is an even number. In this case, its components whose indices are all different are zero. For a component to be non-zero, its indices must be associated in pairs, which results in the presence of Kronecker symbols like $\delta_{i_p \, i_q}$. Moreover, the tensor is completely symmetric. For example, we must have (implicit summation on repeated indices is assumed) 

$$ T_{ij} = A_2 \, \delta_{ij}, ~~{\rm with}~~ A_2 = \di{1\over n}\, \delta_{ij} T_{ij} = \di{1\over n},~~{\rm since} ~~  \delta_{ij} T_{ij} = < \di{\sum^n_{i=1}} \hat{x}^2_i > = <1> =1 $$       

\vv \nin and 

$$ T_{ijk\ell} = A_4 \left[ \delta_{ij}  \delta_{k \ell} +  \delta_{ik}  \delta_{j \ell} +  \delta_{i\ell}  \delta_{jk} \right] ~~ {\rm with}~~ A_4 = \di{1 \over{n(n+2)}} $$ 

\vv \nin It is then obvious that the contraction $T_{i_1\cdots i_k}   \partial_{i_1} \cdots \partial_{i_k}$ 
leads to repeated action of the Laplace operator $\Delta$ : 

$$ T_{ij} \partial_{i} \partial_{j} = A_2 \, \Delta, ~~  T_{ijk\ell} \partial_{i} \partial_{j} \partial_{k} \partial_{\ell} = 3 A_4 \Delta \Delta, \cdots, $$
\beq  T_{i_1, \cdots, i_{2k}} \partial_{i_1} \cdots \partial_{i_{2k}} = \di{ (2k-1)!! \over{\di{\prod^{k-1}_{q=0}}(n+2q) }} \Delta^k \label{projo} \enq 

\vv \nin Hence, said expansion is of the form  

\beq  \bar{u}(P,C) - u(PC) = \di{\sum^\infty_{k=0}} \,C_{2k} \,b^{2(k+1)}\, \Delta^{(k+1)} u (PC) ,~~ {\rm with} \label{expan} \enq 
$$ C_{2k} = \di{1\over{2^{k+1} \,(k+1)! \,\di{\prod^k_{q=0}}(n+2q) }},~~{\rm i.e.} $$ 
$$ C_0 = \di{1\over{2n}},~~ C_2 = \di{ 1 \over{8n (n+2)}} ,~~~ C_4 = \di{1\over{48n (n+2)(n+4)}}, \cdots $$

\vv \nin Both Newton's shell theorem and Gurzadyan's theorem are transposable to $n$ dimensions, provided that the Newton's potential be considered as $1/r^{n-2}$ : 

\vv \nin $\bullet$ the Laplace equation for a spherical symmetric function (and for $r \neq 0$) has, up to a multiplicative constant, the unique non-constant solution $u(r) = U_N(r) = 1/r^{n-2}$ giving $\bar{u}(C) = u(C)$ ; note that this last relation is independent of the closed surface surrounding $C$ because the field $\nabla U_N$ satisfies Gauss's theorem adapted to the $n$ dimensions ;  

\vv \nin $\bullet$ from Eq. (\ref{expan}) for central potentials, the condition $\nabla [\bar{u} - u ] = {\bms 0}$ (independently of $b$) of the Gurzadyan's theorem is equivalent to $\nabla \Delta u = {\bms 0}$ and this last equation has the unique non-constant solution  

$$ u(r) = \di{K_1 \over r^{n-2}} + K_2 r^2 $$ 

\vv \nin The extension of both previous theorems, that requires $\bar{u} - w(b) u  = 0 $  independently of $b$, is also transposable to $n$ dimensions, see \cite{TMP}. From Eq. (\ref{expan}), the condition is equivalent to     
 
$$  u(PC) \left[ w(b) -1 \right] = \di{\sum^\infty_{k=0}} \,C_{2k} \,b^{2(k+1)}\, \Delta^{(k+1)} u (PC) $$ 

\vv \nin and, $PC=a$ and $b$ being independent, this equation can be satisfied if and only if $\Delta u$ and $u$ are proportional, i.e. when $u$ is a solution of the equation 

\beq  \Delta u (a)  = u^{\prime \prime} + \di{{n-1}\over a}\,u^\prime  = c_2 \, u(a) \label{helm} \enq  

\vv \nin $c_2$ being a real constant (i.e. independent of $a$ and $b$). By setting $u = a^{\frac{1}{2} -n} B $ and considering $B$ as a function of the new variable $z$ such that $z^2 = - c_2 a^2$, we find that $B(z)$ must verify the equation  

\beq  z^2 \di{{d^2 B}\over{dz^2}} + z \di{{d B}\over{dz}} + B \left[ z^2 - \nu^2\right] = 0,~~{\rm with}~~\nu = \di{n \over 2} -1 \label{bess} \enq  

\vv \nin which is that of Bessel functions of order $\nu$, see \cite{JEL}. Note that $z$ is real or complex according to the sign of $c_2$. Since $\nu$ is non-integer, the general solution of Eq. (\ref{bess}) is a linear combination of the two Bessel functions $J_\nu(z)$ and $J_{-\nu}(z)$.  

\vv \nin As for $w(b)$, it is expressed as 

$$ w(b) = 1 + \di{\sum^\infty_{k=0}} \,C_{2k} \,b^{2(k+1)} c^{k+1}_2 $$

\vv \nin It is easy to check that for $n=3$, the expressions of $u$ and $w$ are those already obtained in Appendix A.  
\end{document}